\documentclass[a4paper]{VisionStyle}
\usepackage{epsfig}

\def\gsim{\mathrel{\hbox{\rlap{\hbox{\lower4pt\hbox{$\sim$}}}\hbox{$>$}}}}
\def\lsim{\mathrel{\hbox{\rlap{\hbox{\lower4pt\hbox{$\sim$}}}\hbox{$<$}}}}
\def\chandra    {{\em Chandra}\/}
\def\rosat      {{\em ROSAT}\/}
\def\degd       {$^{\circ}\!$}
\def\second     {{\prime\prime}}
\def\mnras        {{MNRAS}\/}
\def\apj        {{ApJ}\/}
\def\apjl        {{ApJ}\/}
\def\aaps        {{A\&AS}\/}
\def\keV{ke\kern-0.05emV}

\begin{document}

\title{Development of Hydrodynamic Instability in the Intergalactic 
Medium of the Merging Cluster of Galaxies A3667}

\author{Pasquale\ Mazzotta\inst{1,2}, Alexey\ Vikhlinin\inst{1},
 Roberto\ Fusco-Femiano \inst{3}, Maxim\ Markevich\inst{1}}

\institute{Harvard-Smithsonian Center for Astrophysics, 60 Garden St.,
Cambridge, MA 02138; mazzotta@cfa.harvard.edu
\and 
Department of Physics, University of Durham,
South Road, Durham DH1 3LE
\and
Istituto Astrofisica Spaziale, Area CNR Tor Vergata,
via del Fosso del Cavaliere, 00133 Roma (Italy)}

\maketitle 

\begin{abstract}

A3667, a spectacular merger cluster, was observed by \chandra ~ twice.
In this paper we review the main results of the analysis of these observations.
In particular we show evidence for the presence in the cluster of  a
 300~kpc Kelvin-Helmholtz hydrodynamic 
instability. We discuss the development of such  instability and
the structure of the  
intracluster magnetic filed in light of a self-consistent cluster 
dynamical model.
 

\end{abstract}

\section{Introduction}

The central region of A3667, a nearby, hot merging cluster (Markevitch,
  Sarazin, \& Vikhlinin 1999), was observed  twice by \chandra
~ in  Sept 22,  1999 and in Sept 9, 2000. 
The analysis of the
first observation by Vikhlinin, Markevitch \& Murray (2001a,b) reveals
the presence of a prominent 500~kpc-long density
discontinuity (``cold front'') in the cluster atmosphere.
This  allowed the authors to draw a cluster dynamical model 
and to put important constraining on the structure of the  
intracluster magnetic field.
A subsequent joint analysis of the two
 \chandra ~ pointings of A3667 by Mazzotta, 
Fusco-Femiano, Vikhlinin (2002) revealed the presence of 
two new interesting filamentary structures in the cluster 
intergalactic medium. These structures are interpreted as 
a well-developed 300~kpc Kelvin-Helmholtz instability.

Here we review the main results of the previous work.
The structure of the paper is as follow.
In paragraph 2 we present the X-ray image and the temperature map 
of the cluster.
 In paragraph 3 we show the properties of the observed features by 
studying their surface brightness and temperature profiles.
In paragraph 4 we discuss the data. 
We start by  presenting the cluster dynamical model in paragraph 4.1. 
Then in paragraph 4.2 we show that,
in a narrow region close to the cold front,
 the data require the presence a well structured
magnetic field. 
We discuss this aspect and we show that the proposed model naturally accounts 
for it. We conclude  by discussing the implications of the proposed model 
for the development of hydrodynamic instabilities.

In the following we skip the details of the data analysis and 
we refer to the original papers  (Vikhlinin et al. 2001a, 
Mazzotta et al. 2002).
We use $H_0=50$~km~s$^{-1}$~Mpc$^{-1}$ and $q_0=0.5$, which corresponds to
the linear scale of 1.46~kpc/arcsec at the cluster redshift $z=0.055$.

\begin{figure}[ht]
  \begin{center}
    \epsfig{file=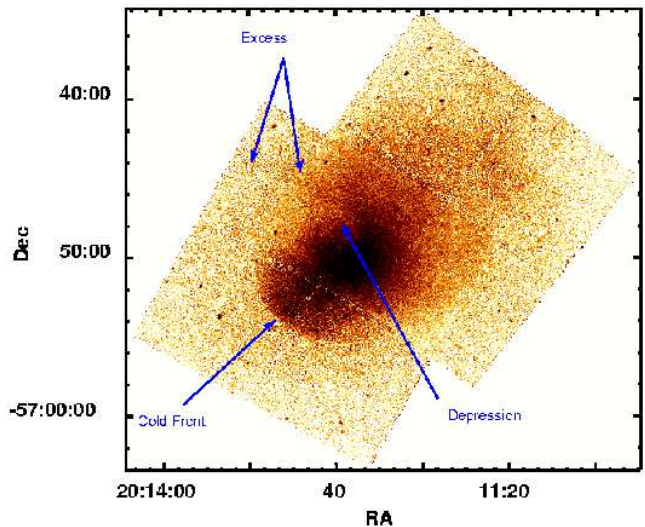, width=8.5cm}
  \end{center}
\caption{Count rates image in the $0.7-4$~keV band binned by $4^\second$.
The arrows indicate the prominent X-ray features: the cold front,
a filamentary arc-like surface brightness excess 
extending toward the east, and a filamentary arc-like 
surface brightness depression extending toward the west.}   
\label{pmazzotta-B1_fig:fig1}
\end{figure}

\begin{figure}[ht]
  \begin{center}
    \epsfig{file=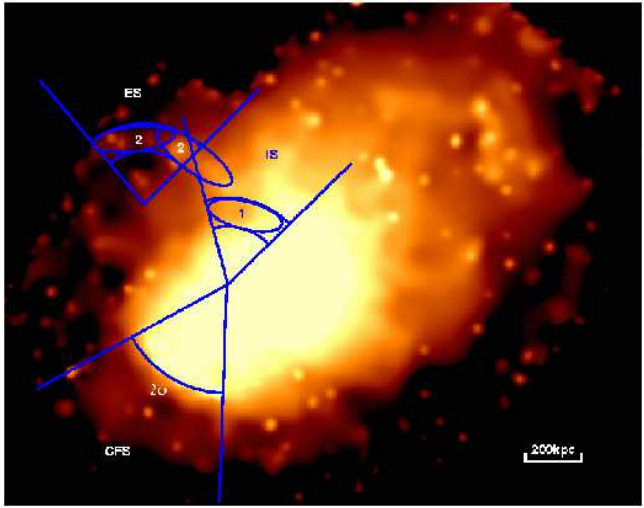, width=8.5cm}
  \end{center}
\caption{Adaptively smoothed \chandra ~ image of A3667. 
The six lines identify the Cold Front Sector (CFS),
the External Sector (ES), and the Internal Sector (IS)
used to derive the surface brightness and temperature profiles reported 
in Fig.~\ref{pmazzotta-B1_fig:fig4} and Fig.~\ref{pmazzotta-B1_fig:fig5} .
The respective position angles are: from 110\degd ~ to 170\degd,
from -15\degd ~ to 15\degd , and from -15\degd ~ to 40\degd ~
 for CFS, IE, and ES, respectively 
(angles are measured from North through East). 
Regions 1 and 2 identify the X-ray  depression and excess, respectively.}
\label{pmazzotta-B1_fig:fig2}
\end{figure}

\section{X-ray image and temperature map}

In  Fig.~\ref{pmazzotta-B1_fig:fig1} we show the 
background subtracted and vignetting corrected   image 
of the central region of A3667 obtained combining the two 
\chandra~ pointings.
The image is extracted in the 0.7-4~keV band
to minimize the relative contribution of the cosmic background and 
thereby to maximize the signal-to-noise ratio.  
The X-ray image reveals three prominent features:

i) a sharp surface brightness edge spanning 500~kpc to the South-East; 

ii) a $\approx 300$~kpc-long and $\approx 90$~kpc-wide
 filamentary X-ray excess extending toward the east to the chip boundary;

iii) a $\approx 200$~kpc-long and $\approx 75$~kpc-wide
 filamentary X-ray depression that develops toward the west inside the
cluster center.

All the three features are also evident in the smoothed image 
reported in Fig.~\ref{pmazzotta-B1_fig:fig2}.
To derive the surface brightness and temperature
profiles  for each feature (see \S~\ref{pmazzotta-B1_s:s3}, below), 
three different sectors have been identified.
Each sector originate near the curvature center of 
the corresponding feature and contains the feature itself.

For convenience, in the following, we refer 
to the sector containing the surface brightness edge as the 
cold front  sector (CFS) and to the sectors containing the excess
and the depression as the External Sector (ES) and the Internal Sector (IS),
respectively.  
To facilitate the identification of the features in
the temperature map we traced  the X-ray depression and excess 
using  two ellipsoid regions indicated by the
cardinal numbers 1 and 2, respectively. 

\begin{figure}[ht]
  \begin{center}
    \epsfig{file=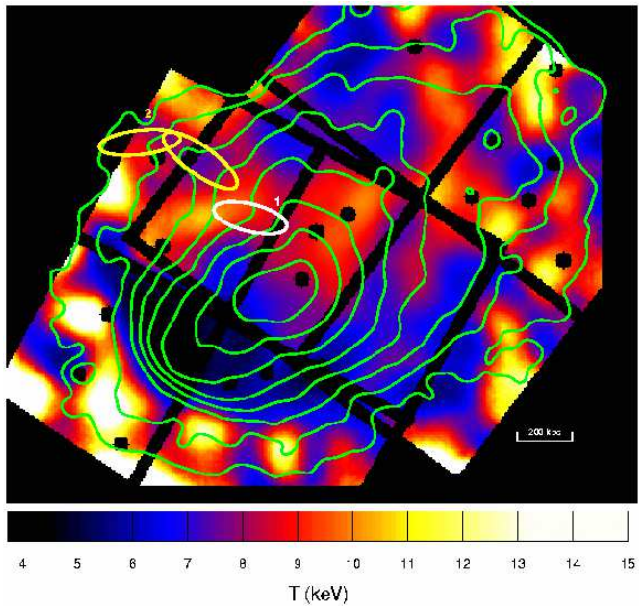, width=8.5cm}
  \end{center}
\caption{Temperature map  with overlaid ACIS-I X-ray
surface brightness contours (spaced by a factor of $\sqrt{2}$) in the
0.7-4~keV energy band after adaptive smoothing.   The black cut-out regions identify
  the point sources that were masked out.
The regions indicated by cardinal numbers 1 and 2 are  the same shown in 
Fig.~\ref{pmazzotta-B1_fig:fig2}.}
\label{pmazzotta-B1_fig:fig3}
\end{figure}

In Fig.~\ref{pmazzotta-B1_fig:fig3} we report the 
temperature map with  overlaid X-ray
surface brightness contours. 
The temperature map shows a quite complex structure far from being isothermal.
Moreover, we notice that:

a) the gas in the brighter side of the edge  is cooler than the 
gas in the fainter side; 

b) the temperature of the X-ray excess (region 2) 
is lower than in the nearby regions; 

c) the  temperature of the X-ray depression (region 1) 
is higher than in the nearby regions.

We also note that, in the overlapping region, 
the Chandra temperature map qualitatively agrees  
with the coarser ASCA temperature map shown by 
Markevitch et al. (1999).

\section{Surface brightness and temperature profiles}\label{pmazzotta-B1_s:s3}

In the following we study the physical properties of the features by extracting the surface brightness and the temperature profiles from the sectors 
defined in the previous paragraph.

\subsection{Cold front}

\begin{figure*}[ht]
  \begin{center}
\epsfig{file=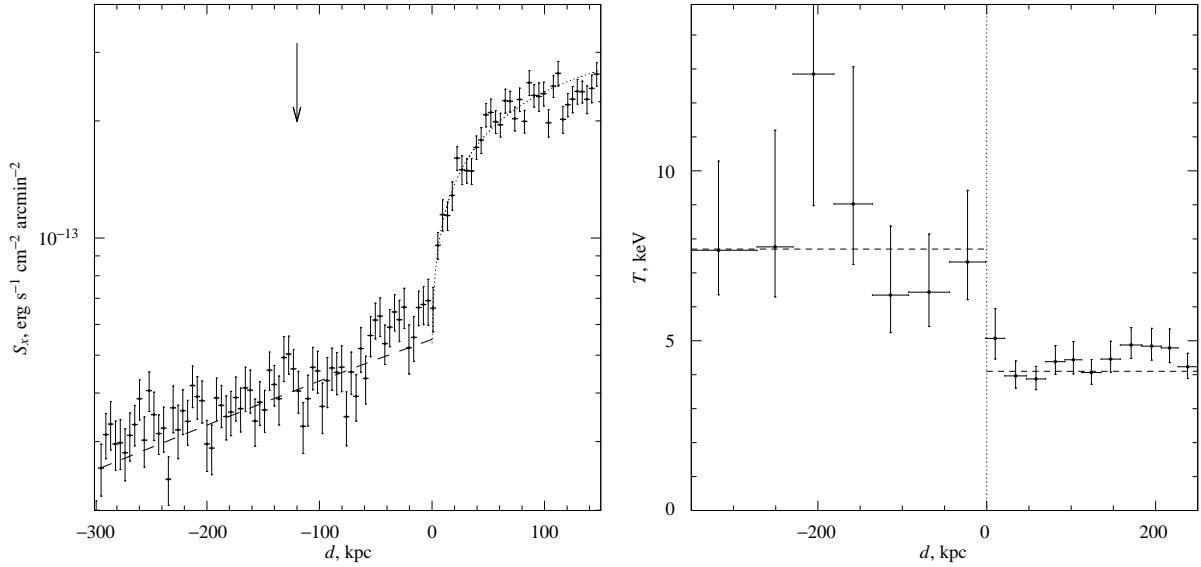, width=16cm}
  \end{center}
  \caption{ \emph{(a)}: X-ray surface brightness profile (expressed in the
    units of energy flux in the 0.5--2~\keV\ band) across the cold front.
    The profile was measured in the CFS show in Fig.~\ref{pmazzotta-B1_fig:fig2}.
The distance is measured relative
    to the front position. 
    Temperature profile across the front measured in the same sector as the
    surface brightness profile. The dashed lines show the mean gas
    temperature inside and outside the front.}
\label{pmazzotta-B1_fig:fig4}
\end{figure*}

In  Fig.~\ref{pmazzotta-B1_fig:fig4}a we report the 
 surface 
brightness profile extracted from the CFS.
The figure clearly shows that the surface brightness profile increases 
sharply by a factor 2 within a very 
small region of 40-80~kpc. 
This strongly indicates that the gas density profile is  
discontinuous. Vikhlinin et al. (2001a), modeled the observed surface brightness profile 
with a denser spheroidal gas cloud embedded in a more rarefied 
cluster-like atmosphere
described by a $\beta$-profile. They find, indeed, that the surface brightness 
profile is well fitted by this model and  that the gas 
discontinuity  at the edge is a factor $3.9\pm0.8$.   

In Fig.~\ref{pmazzotta-B1_fig:fig4}b we report the temperature 
profile extracted from the same region. It
shows a clear temperature jump at the surface brightness edge. 
Indeed,
as we cross the edge from the outside, the temperature changes 
abruptly from approximately 8~keV to 4~keV. 
We stress the fact that although gas density discontinuities can be produced 
by shock fronts, the observed temperature jump is not consistent with this 
hypothesis. In fact, the observed temperature jump goes in the opposite
 direction with respect to what expected in a shock front. 
Thus, the feature we are observing is produced by a new phenomenon
which has been called ``Cold Front'' (Vikhlinin et al. 2001a).

The derived density and temperature jumps lead to a significant pressure jump.
In particular one can calculate that  the pressure outside 
the edge is approximately a factor 2 lower than the pressure 
of the gas just inside the edge.

\subsection{X-ray filaments}

\begin{figure*}[ht]
  \begin{center}
    \epsfig{file=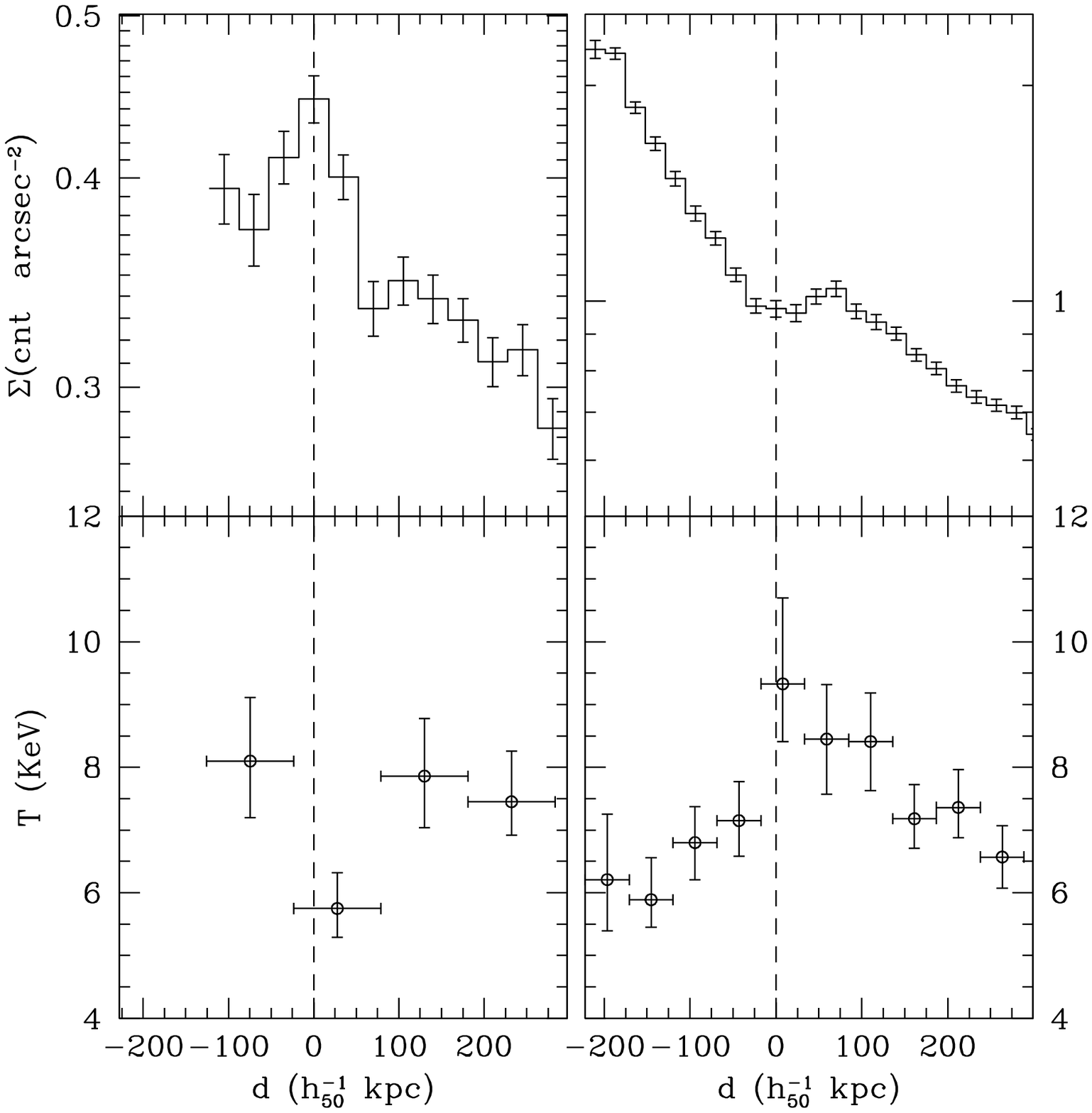, width=16cm}
  \end{center}
\caption{X-ray surface brightness (upper panels) and temperature  (lower panels)
profiles from the ES (left column) and IS (right column), 
defined in Fig.~\ref{pmazzotta-B1_fig:fig2}. 
The dashed line in the left column indicates the relative maximum in the 
ES X-ray surface brightness profile, while 
the dashed line in the right column indicates the relative minimum in the 
IS X-ray surface brightness profile.
The x-axis    indicate the distance in kpc from the surface brightness profile stationary point 
of the corresponding sector.
Error bars are at $68\%$ confidence level.}
\label{pmazzotta-B1_fig:fig5}
\end{figure*}

For each of the two filamentary structures,
we extracted both the surface brightness and the temperature profiles.
The profiles are shown in the left and the right column of
Fig.~\ref{pmazzotta-B1_fig:fig5} for the ES and IS, respectively.  
  The ES profiles
show that the filamentary excess produces a significant enhancement of the
surface brightness profile and corresponds to a decrement in the temperature
profile.  On the other hand, the IS profiles shows that the filamentary
depression produces a significant decrement of the surface brightness
profile corresponding to an increment in the temperature profile. 
 This suggests that the excess is a colder, denser filamentary gas structure
embedded in the more diffuse and hotter external cluster atmosphere while
the depression is a hotter, rarefied filament of gas embedded in the denser
and colder cluster center. 
We also notice that, while the temperature of the X-ray filamentary 
excess is consistent with the temperature of gas surrounding 
the filamentary depression,  the temperature of the X-ray filamentary 
depression is consistent with the temperature of gas surrounding 
the filamentary excess.
The observed features suggest that some dense cold gas is ``striped out'' into the hotter cluster atmosphere as well as some rarefied hot gas from larger radii penetrates into the cluster core.

\begin{figure*}[ht]
  \begin{center}
\epsfig{file=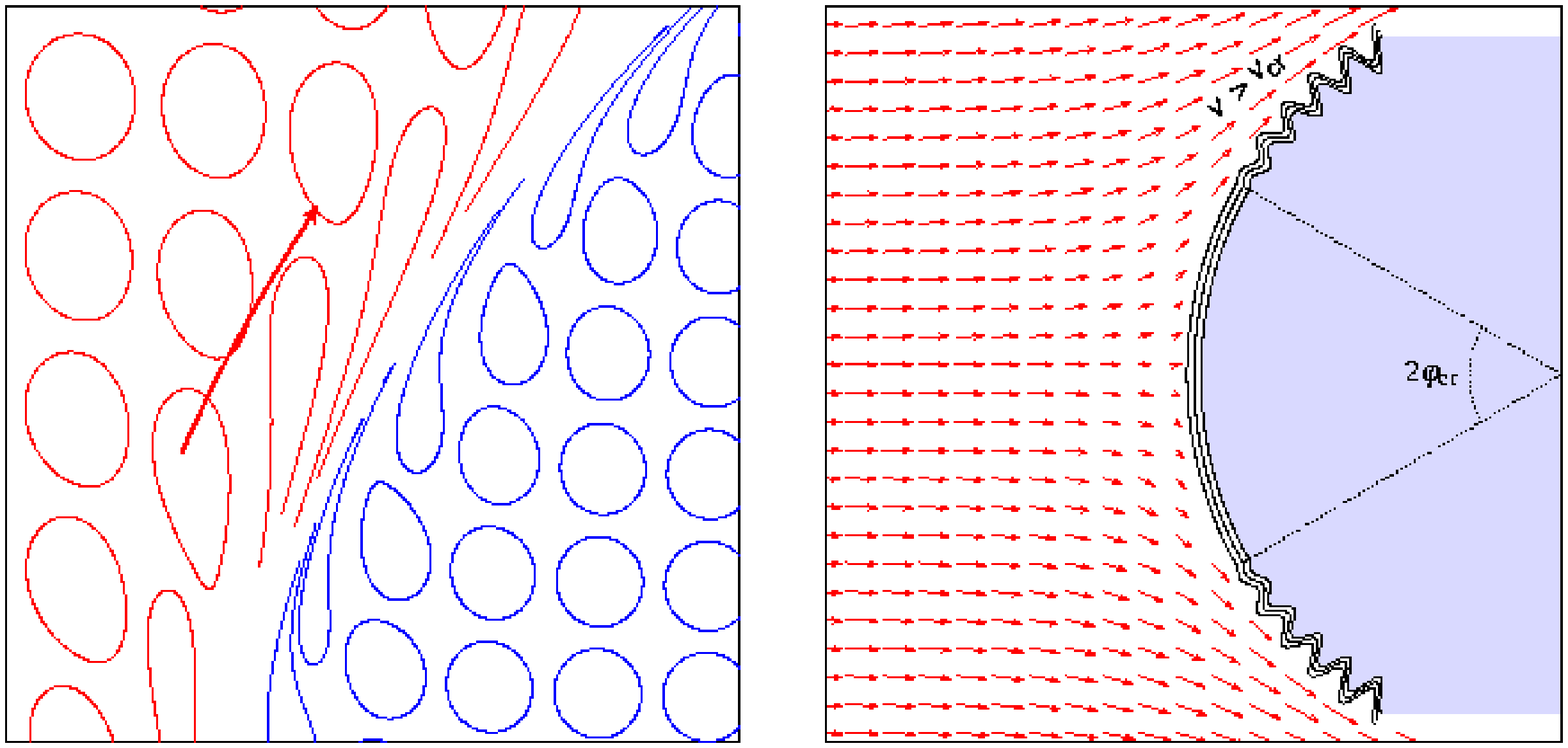, width=16cm}
  \end{center}
  \caption{\emph{(a)} Illustration of the formation of the magnetic
    layer near the front surface. The initially tangled magnetic lines in
    the ambient hot gas (red) are stretched along the surface because of
    tangential motion of the gas. The magnetic lines inside the front are
    stretched because, in the absence of complete magnetic isolation, the
    cool gas experiences stripping. This process can form a narrow layer in
    which the magnetic field is parallel to the front surface. Such a layer
    would stop the transport processes across the front, as well as further
    stripping of the cool gas. \emph{(b)} The interface between the cool and
    hot gas is subject to the Kelvin-Helmholtz instability. The magnetic
    layer can suppress this instability in the region where the tangential
    velocity is smaller that a critical value $V_{\rm cr}$. The velocity
    field shown for illustration corresponds to the flow of incompressible
    fluid around a sphere.}
\label{pmazzotta-B1_fig:fig6}
\end{figure*}

\section{Discussion}

In the previous paragraph we showed evidence for the presence 
of a cold front  and two filamentary structures in the 
atmosphere of A3667. These features are particularly interesting as them allow
us to understand the cluster dynamic 
and the  undergoing intracluster
physical processes.
In the following we address some of these topics  discussing a self-consistent 
dynamic model that accounts for all the properties of the observed features.

\subsection{Cold front and cluster dynamics}

A3667 is a know merger cluster.
The  \rosat ~ image show that, at any radius, 
the cluster X-ray emission is elongated in the  
direction  from North-West to South-East 
(see e.g. Markevitch et al. 1999). 
From Fig.~\ref{pmazzotta-B1_fig:fig1} 
  we can see that the elongation of the diffuse X-ray emission
 is  almost perpendicular to the observed cold front.
It appears natural, then, to conclude that  the cold front is 
simply produced by the motion of one merging subclump.
In this scenario the observed cold front would be the contact surface 
of a dense cold gas cloud  that moves from Noth-West to South-East into the 
hotter rarefied cluster atmosphere (Vikhlinin et al. 2001a). 
The pressure  inside the cloud  would then be higher
than the gas outside as it is subject to ram pressure in addition 
to the  thermal pressure of the cluster atmosphere.
Using  the pressure jump at the front, Vikhlinin et al. (2001a)  
determined the motion speed and found the near-sonic velocity 
$M\equiv v/v_s \gsim 1$
(here $v_s$ is the speed of sound in the hot
ambient gas).

Fig.~\ref{pmazzotta-B1_fig:fig4} shows that the surface brightness 
of the front is particularly sharp. This gives us two other important 
pieces of information:

a) the gas cloud is moving in plane 
close to the plane of the sky. In fact, 
because of  projection effect, a large departure from this situation would
make the front to appear smoother 
(Mazzotta et al 2001).
This is also supported by the fact that the difference in line-of-sight 
 velocity  of the two dominant galaxies is only 120~km~s$^{-1}$ 
(Katgert et al. 1998), indeed much  
smaller than the estimated speed of the cloud;

b) the front width  is particularly small.
Vikhlinin et al. (2001a)  measured the front width and found 
that it  is $<5$~kpc, indeed
smaller than the Coulomb mean free path of the intra-cluster gas. 
This means that, at least in the direction perpendicular to the front, 
transport processes, and thus 
thermal conduction, are highly suppressed in all the narrow region 
containing the front (see also Ettori and Fabian 2001).
Most likely such a suppression is induced by the presence of a
magnetic field parallel to the front as we are going to discuss in the 
next paragraph.

\subsection {Magnetic field structure}

Magnetic fields in a  highly ionized plasma act to 
suppress  transport processes in  the direction 
perpendicular to the magnetic field  lines.
In order for transport processes to be effectively suppressed 
on a scale comparable to the front length
[as required by point b) in paragraph 4.1] we should either have that:

i) the magnetic filed lines are  
highly tangled (see e.g. Rechester \& Rosenbhluth 1978); 

ii) there is a large-scale magnetic field 
perpendicular to the direction in 
which diffusion is to occur.

Although  hypothesis i) has been invoked by many authors
to justify the needs for  suppression of   thermal conduction in clusters,
in a recent paper Narayan \& Medvedev 
(2001) show that, if the medium is turbulent,  this is not  longer true. 
Indeed, thermal conduction in a highly tangled magnetized turbulent
plasma is almost as efficient as in a non magnetized plasma.
Therefore, in the case of A3667, hypothesis ii) appears to be more likely
to explain the suppression of transport processes at the front.
Vikhlinin et al. (2001b)
suggested that, in a narrow  boundary region 
between the two moving gas layers,
 the required magnetic field may naturally arise 
 through a process schematically shown in Fig.~\ref{pmazzotta-B1_fig:fig6}.
During the gas cloud motion, the surrounding  gas flows around the cold cloud.
The inflowing gas slows down near the stagnation 
point at the leading edge of the sphere but then reaccelerates to 
high velocities as it is squeezed to
the sides by new portions of the inflowing gas 
(see Fig.~\ref{pmazzotta-B1_fig:fig6}b).
The gas is stripped from the surface of the cloud and the magnetic 
field lines, which are frozen into the gas and initially tangled, 
 stretch along the front (see Fig.~\ref{pmazzotta-B1_fig:fig6}a).
When a layer with a parallel magnetic field forms, it prevents the
 further stripping of the cold gas and also stop the 
transport processes across the front.

As we will see  in the next session,
such a layer with a parallel magnetic field acts, also,
to suppress the  development of hydrodynamic instabilities

\subsection{Hydrodynamic instability} 

One interesting  issue  of the proposed cluster dynamical model is that 
the interface between the tangentially moving gas layers
must develop both Rayleigh-Taylor (R-T) and Kelvin-Helmholtz (K-H)
instabilities (see Inogamov, 1999 for a review). 

The R-T instability develops at the interface  between the two fluids when the
rarefied fluid accelerates the denser one.  If, as it is reasonable to
assume, the gas cloud moves together with its own dark matter halo 
(Vikhlinin \& Markevitch 2002, in preparation)
the gas cloud is stabilized against the R-T instability by 
gravity, so that the development of R-T
instability is quite unlikely.  In contrast the K-H instability is related
to the shearing motion at the boundary between two fluids and is expected to
develop along the lateral boundaries of the cloud\footnote{
Long-wavelength  K-H perturbations are also stabilized by a 
gravitational field perpendicular to the interface. 
However, Vikhlinin (2001b) found that in the case of A3667 only 
 $\lambda > 4500$~kpc modes are stable.
Therefore gravity is unimportant for the scales considered here.}

The wavevector  of the fastest
growing mode, $\lambda$, is parallel to the flow and its growing 
time $\tau$ is given by
solving the dispersion equation (see e.g. Miles 1958).
Vikhlinin et al.\ (2001b) computed the K-H instability growth time for the
flow near the cold front in A3667. 
As the only perturbations relevant for our discussion are the ones that 
grow on time scales shorter than the cluster core
passage time, $t_{cross}=L/v_{cross}$ (here $L$ is the cluster size and
$v_{cross}$ is the motion speed), we  derived the ratio between 
these two times\footnote{This equation reproduces eq.(4)
  from Vikhlinin et al.\ (2001b), corrected for an algebraic error which
  resulted in overestimation of $\tau$ by a factor of $4\pi^2$. Note that
  the correct equation does not affect any results presented in Vikhlinin et
  al., and indeed, strengthens their arguments.}:

\vskip 0.5truecm
\begin{equation}
  \frac{t_{cross}}{\tau}=3.3 {L \over \lambda} \sin \varphi,
  \label{pmazzotta-B1_eq:1}
\end{equation}
\vskip 0.5truecm

\noindent
where $\varphi$ is the angle between the perturbation and the leading edge
of the moving cloud.

From Eq~1. we see that:

a) in absence of stabilizing factors, perturbations grown on all scales;

b) small scale perturbations grow faster than the larger scale ones;

c) for a fixed wavelength the growth time is shorter at larger angles
$\varphi$ reaching its minimum at $\varphi=90$\degd .  This is a direct
consequence of the fact that the speed of the external fluid increases with
$\varphi$ being maximum at $\varphi=90$\degd .

Using Eq.~1 and assuming  
$L\approx 1$~Mpc,
 we find that, already  at very small angles
($\varphi \gsim 10-50$~arcmin),  the growth time of a 
$10$~kpc  perturbation  is much shorter
 than the cluster passage time.
 Then, because of points a) and b) above, 
  there should develop a turbulent layer which
 would smear out the cold front by at least $\sim 10$~kpc at angles 
 $\varphi \gsim 10-50$~arcmin.  
As evident from Fig.~\ref{pmazzotta-B1_fig:fig2} 
such a smearing is not present in the \chandra ~ image, at least  
within  the $\pm30^\circ$ sector of the leading
 edge of the cold front (CFS in Fig.~\ref{pmazzotta-B1_fig:fig2}).
The same figure shows front smearing only at angles  $\varphi > 30^\circ$.
This behavior is consistent with the 
 proposed magnetic field configuration, indeed. In fact,
beside suppressing the transport processes,
 the surface tension of the amplified magnetic 
 field acts to stabilize the development of the K-H instability.
As the fluid speed is lower, the K-H instability is effectively stabilized 
only inside the  $\pm30$\degd ~ sector.
At larger angles  the flow speed gets higher and 
 the magnetic field  surface tension becomes insufficient to 
stabilize the front. 
This argument allowed Vikhlinin et al.\ (2001b) to constraining 
the strength of the magnetic field finding $7\mu$G$<B<16\mu$G.

Outside the $\pm30^\circ$
sector, the development of the K-H instability is unaffected by the magnetic
field, and therefore the perturbation growth time is given by
eq.~\ref{pmazzotta-B1_eq:1}.
In this region the evolution of the front 
is  as follow.
At first,  when the interface is still a
discontinuity, the front develops small scale instabilities.
Their growth effectively widens the interface, which becomes a
turbulent layer of finite width.  At this
point the perturbations on scales smaller than the evolving width of the
front are damped (they becomes unobservable as individual
structures) while perturbations on a larger scale start to
grow (thus they may appear as distinct  structures) 
(see e.g. Esch 1957). As this process
continues, it is expected that the remaining observable wavelengths are the
ones with a growth time comparable to the cloud crossing time, namely
$t_{cross}/\tau\sim 1-10$.

In this scenario the  excess and the depression filamentary 
structures observed on the side
of the moving subclump may naturally be interpreted as the result of  
the development of a large scale K-H instability.
We notice, in fact, that the structures lie, at $\varphi \approx 90$\degd ~
which,  because of point  c) above, is indeed a privileged 
point for the growth of the perturbations.
Furthermore, if we assume that $\lambda=300$~kpc and
$L=1$~Mpc, the growing time for this perturbation 
is such that $t_{cross}/\tau\approx 10$. This means
that the perturbation is just now entering the strong non-linear regime,
 consistently with what we observe.

The discovery of this instability 
is particularly important to fully understand the 
physics of mergers.
One interesting aspect, for example, is that  although
the cold front prevents the gasses of the merging objects to mix along the
direction of motion, strong turbulent mixing processes, on scales comparable
to the size of the merging subclump, may occur on the sides.
We notice that this process may affect 
the thermal evolution of the merging subclump. 
It, in fact, clearly favorites  the deposition of 
rarefied hot gas right in the center of the cluster
(see Fig.~\ref{pmazzotta-B1_fig:fig1}). 
At the same time, it  pulls out some dense cold gas from  the same 
 region. 
If efficient, this process may have important consequences 
for the development and/or evolution  of a central  cooling flow.

\section{Conclusion}

We reviewed the main results obtained from the analysis 
of  two \chandra ~ observations of the central
region of A3667.  
The cluster hosts a spectacular and well defined cold front.
Moreover, it shows 
 two arc-like adjacent filamentary structures: one,
extending from the colder subcluster toward the cluster
outskirts, appears as a dense structure embedded in the less dense cluster
atmosphere, and the other, extending inside the
  subcluster, appears as a rarefied structure
embedded in the denser cluster core.

We suggest that the filamentary structures  represent the first evidence for the
development of a large scale hydrodynamic instability in the cluster
atmosphere. 
We discussed this problematic in light of a self-consistent  
cluster dynamical model.

\begin{acknowledgements}
P.M. thanks the conference organization committee for waiving his registration 
fee for the conference.
Support for this study was provided
by NASA contract NAS8-39073, grant NAG 5-9217, and by the Smithsonian
Institution. 

\end{acknowledgements}

\end{document}